# Remote Mentoring Young Females in STEM through MAGIC

Ritu Khare[1], Esha Sahai[1,2], Ira Pramanick[1,3]
[1]GetMAGIC, [2]Massachusetts Institute of Technology, [3]Google
ritu.khare@nih.gov, esha@sloan.mit.edu, ira@getmagic.org

**Abstract**

*The limited representation of women in STEM workforce is a concerning national issue. It has been found that the gender stratification is not due to the lack of talent amongst young females, but due to the lack of access to female role models. To this end, "remote mentoring" is an effective way to offer nation-wide personalized STEM mentoring to young females from all segments of the society. In this paper, we introduce MAGIC, an organization dedicated to mentoring young females in STEM through remote methods. We conduct a retrospective study of MAGIC's formative years and present our experience in remotely establishing 23 highly tailored mentor-mentee pairs. We provide several key findings on STEM remote mentoring, such as popular communication tools, frequently sought STEM skills among girls, and projects that could be accomplished through remote mentoring. Furthermore, we present key challenges faced by mentors and mentees, notable outcomes, and lessons learnt about remote mentoring.*

## 1. Introduction

Despite nation-wide efforts to improve women empowerment and equality, women represent only 24% of the science, technology, engineering, and mathematics (STEM) workforce in the United States [1]. This gender stratification and limited participation is not due to the lack of talent amongst girls, but due to the discouraging societal attitudes and unique challenges a girl faces in the formative stages of her life [2, 3]. It is thus imperative to establish a nation-wide mentoring system that engages, motivates, and inspires [4] young females toward STEM subjects in a personalized manner. To this end, we systematically explore a "remote mentoring" approach to STEM mentoring, wherein the "mentee", the girl receives mentoring services using telecommunications and internet technologies. Due to the ubiquitous nature of technology, remote mentoring offers the following advantages over traditional face-to-face mentoring:

- Accessibility: Remote mentoring allows reaching out to girls located in those remote regions that are technologically advanced but lack access to women role models [5, 6] and mentoring resources.
- Personalization: A successful mentor-mentee relationship is highly tailored; both parties should be able to establish a common ground in terms of interests and values, and the mentor must possess certain qualities that the mentee aspires to emulate. The odds of finding a "matching" mentor drastically improve if geographic location is not a constraint as in the case of remote mentoring.
- Efficiency: Traditional mentoring is expensive; it involves scheduling face-to-face meetings, planning and reserving a suitable venue, and commuting to the venue. On the other hand, remote mentoring is location independent, and hence it eliminates the commute time and cost involved in traditional mentoring. Additionally, remote mentoring offers flexibility in scheduling meetings.

Therefore, we believe that remote mentoring is the only way to scale nationwide and offer personalized STEM mentoring to girls from all segments of the society.

In this paper, we introduce our remote mentoring based organization, Get *More Active Girls in Computing* (GetMAGIC or MAGIC for short)[7], founded in 2008. MAGIC (http://getmagic.org/) is a non-profit organization dedicated to encouraging young females to pursue STEM academic degrees and careers. MAGIC acts as a nationwide matching bridge between girls who are interested in STEM topics and women with successful technology and computing careers. In this study, we present the first data on remote mentoring of young females in STEM. A five-year retrospective statistical analysis of remote mentoring interactions indicates that it is possible to achieve personalization and accessibility through remote mentoring. We have established mentoring relationships across at least seven US states and mentored 23 girls from three different types of schools having different levels of expectations from the mentoring experience. The girls were given opportunities to develop a diverse set of STEM skills and work on a variety of STEM projects through remote mentoring. Some key results include:

- Skype and Google Hangout are the most popular communication tools used by 39% and 21% of the mentor-mentee pairs, respectively.
- While MAGIC aims to mentor girls in a large variety of STEM fields, about 80% of the mentees choose to work on computer programming based projects, involving computation and game design. An explanation to this skewed distribution is that majority of the mentors possess a strong background in computer science.
- Mentees learn four types of STEM skills through remote mentoring: programming languages, (42%), topic-based knowledge (31%), creative tools (18%), and Web development (9%).
- Mentors and mentees perceive different sets of challenges in remote mentoring. While mentees report time management and logistics as main hurdles, mentors identify the lack of face-to-face interaction and STEM concept delivery as main challenges.
- Overall, the most notable impact on the mentees is observed with respect to building skills and confidence, improving scientific visibility, and piquing interest in career growth.

These results collectively help to understand the desirable practices in remote mentoring, assess the interests of young females, recruit suitable mentors, and design remote programs accordingly.

## 2. Related Organizations and Studies

MAGIC is inspired by several organizations that continue to have a significant impact on society by supporting women and minorities. MentorNet[8], founded in 1997, is a STEM e-mentoring organization that aims to increase the representation of women in scientific and technical fields by using a dynamic mentoring network. Iridescent[9], a non-profit organization to encourage girls to pursue technology entrepreneurship, organizes a 12-week program that matches each mentor with multiple mentees to assist them in designing mobile phone applications that are finally pitched to real venture capitalists. Although this program is primarily face-to-face, the organization is exploring web technologies for its expansion. In several other organizations, such as Big Sisters[10], e-mails are an important facet of the mentor-mentee relationship building. Girls Who Code[11] is another recent effort in encouraging girls to pursue computer science using face-to-face teaching and mentoring sessions. Amidst these organizations, MAGIC is unique in two significant ways. First, we mentor teenaged females, who require more consideration, compassion, and caution than adults do. Second, we conduct long-term systematic mentoring instead of unstructured, ad hoc sessions. In addition, the program is highly personalized for each mentee. The mentor facilitates the mentee in discovering her specific interests, learning STEM skills, and gaining hands-on experience.

There are very few data-driven studies on remote mentoring children and teenagers. Brown and Dexter [12] conducted a two-year retrospective study of one-on-one remote mentoring interactions between 300 elementary school students and 100 business professionals. The objective of the program is to assist students to write more and make fewer mistakes, and develop computer skills. Each mentor-mentee pair used e-mail programs and MS Office Word, and conducted weekly writing activities on a variety of topics. The key findings of the study were that successful e-mentoring required enthusiastic and committed mentors, personal connection between mentor and mentee, technical support, a steering committee, and a smaller setup. This in turn would lead to the development of social, communication, and academic writing skills among students.

It has been found that gender determines mentoring outcomes [13], and recently several studies have focused on same sex mentoring based on the traditional face-to-face model:

- Spencer and Liang [5] conducted an analysis of 12 female relationships, focusing on healthy psychological development and emotional intimacy. The pairs were based in the greater Boston area and were established through Big Sisters association. The study conducted qualitative interviews with mentors aged 28-55 and mentees aged 13-17, and empirical analysis of their 2.5 - 11 yr. long relationships. The main findings were that mentees felt engaged, experienced authentic emotional support, developed new skills, gained confidence, and found companions. The study recommends that mentoring programs should prioritize emotional support over instrumental support.
- Tyler-Wood et al. [14] piloted an analysis of an NSF funded five-year mentoring program in environmental science and related careers. The mentors were female high school students, and the mentees were 32 elementary school students from a community in North Texas. Some adult mentors served as core mentors to oversee the program activities. The authors conducted a survey-based study on the short and long term impact of students' perceptions of STEM careers. It was observed that the participants made significantly higher gains in academic scores, and a follow up study with 14 mentees showed positive impacts on their perceptions after six years.
- Khoja et al. [2] conducted a retrospective study on a four-week computer science camp attended by middle school girls. The camp focused on exposing girls to computer science field and changing their perception of computer science. The girls worked on Alice, Lego, and social media projects and filled out a daily evaluation survey. The camp helped the girls to develop better attitudes toward computers and computer scientists, and to develop self-confidence and technical skills.

Additionally, other non-empirical studies provide evidence that mentoring is perceived positively by youth as well as adults[15]. Similar to existing studies, we focus on mentoring young females [5, 15, 16], particularly in STEM [6, 14, 17], using remote technologies [12, 18]. However, our study differs in that it conglomerates young females, STEM, and remote mentoring, and it provides the first empirical data-driven evidence of associated findings.

## 3. Organizational Setting: MAGIC

MAGIC was founded by a small number of highly dedicated women, who are committed to encouraging more girls to consider STEM careers. MAGIC is in its fifth year, and continues to be administrated by a small board of directors who share the same passion for motivating girls into STEM. The organizational structure of GetMAGIC is flat, flexible, and distributed. At the center, MAGIC contains a core team comprised of a handful of members who oversee the day-to-day operations and make strategic decisions. The MAGIC board has six members who meet quarterly to discuss the strategic and long-term growth of the organization, and provide innovative ideas to overcome challenges. Furthermore, MAGIC attributes its success to the mentors who are women volunteers with successful STEM careers. Each mentor is recruited through a rigorous selection process and is responsible to offer committed mentoring services. All mentors meet with each other on a monthly basis to share successes as well as challenges and offer advice to each other.

MAGIC advocates and practices one-on-one mentoring. This allows the mentors to provide the necessary attention required to teach and delve into the nuances of STEM topics, and to tailor the projects and discussions to the mentee's interests. MAGIC establishes each mentor-mentee pair with an objective to maximize successful outcomes. The matching is performed based on several factors including the mentee's interests and maturity level, e.g., the ability to handle remote mentoring, schedules and availability of mentor and mentee, expertise of mentor, and maturity level of mentor especially in case of socio-economically underprivileged mentee. Each mentoring relationship officially lasts for four to seven months and culminates in a final project. At the end, each mentee receives the opportunity to present her project in the final MAGIC meeting held in one of the partner schools. Some noteworthy final projects include personal website development, mobile phone application programming, Google SketchUp based building design, Tic-tac-toe game design, science fair experiment design, etc. On numerous occasions, these projects have inspired mentees to consider STEM as an attractive career option. Furthermore, the mentors find it very rewarding to watch a young female grasp a considerable amount of technical material in a short span of time.

## 4. Methodology

We collected data on remote mentoring over the course of five years (2008-2013). The information about the mentors and the mentess was collected at the time of recruitment and interviewing. The information about the schools was collected when the partnership was established. The information about the mentor-mentee pairs was collected at several points during the MAGIC mentoring cycle where the core team communicated with the mentees and mentors on an individual basis. The information was updated every time the mentee or the mentor chose to inform or consult the core mentors about a challenging issue. All the information was originally collected by the core team in a narrative fashion, and was anonymized before sharing with the key investigators of this study. Overall, we collected data on 23 mentor-mentee pairs, comprising 16 remote mentors and 23 mentees from 12 participating schools, as shown in Table 1.

Table 1. MAGIC Data used for analysis

| Entity | Total | Information |
|---|---|---|
| MAGIC Remote Mentors | 16 | Highest earned academic degree, background, location, motivation to join MAGIC, number of years with MAGIC |
| Participating Schools | 12 | Location, public/private, number of years with MAGIC |
| MAGIC Mentees | 23 | School, grade, year, motivation to join MAGIC |
| MAGIC Pairs (Mentor-Mentee) | 23 | Basis of match, STEM Topics learnt, projects accomplished, communication and technological tools used, challenges faced by mentor and mentee, impact on mentor and mentee |

With this data, we reported and computed several results and findings from MAGIC's formative five years. We provided a summary of the mentors including a quantitative distribution of academic degrees, backgrounds, location, and motivation to offer their services as remote mentors. We also provided the distribution of school types, mentees' grades, and mentee's motivation to join MAGIC. We also analyzed the evolution of MAGIC workforce over the past few years. More importantly, we did an in-depth analysis of the MAGIC pairs' information. We identified the top tool choices amongst mentors and mentees, most popular STEM topics learnt during remote mentoring, and some sample projects that could be accomplished remotely. We used the traditional qualitative analysis technique of coding to identify themes from narrative texts to identify the top challenges and impact on mentors and mentees. Finally, we synthesized the results to draw collective implications for remote mentoring.

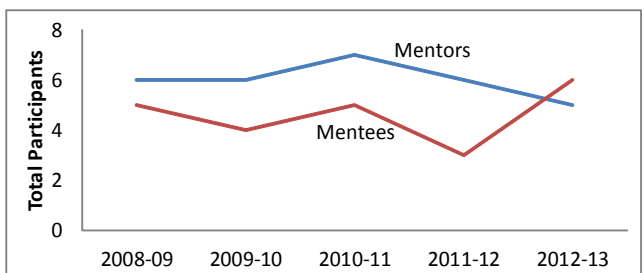

Figure 1 MAGIC Workforce Evolution for Remote Mentoring

## 5. Results
### 5.1 MAGIC Mentors and Mentees

In the past five years, MAGIC facilitated 16 women in offering their remote mentoring services to 23 young females. Figure 1 shows the evolution of MAGIC in the past few years. The red line represents the participating mentees, and the blue line represents all the recruited and active mentors in a given year. The mentors are spread across seven states in the US, with majority of them located in California and Massachusetts, as shown in Figure 2a. The highest academic degrees, finished or in progress, are nine doctorates, five masters, and two bachelors. The mentors come from a variety of STEM backgrounds (see Figure 2b) including computer science, information science, information technology, ocean engineering, astronomy, human computer interaction, robotics, and economics. During recruitment, the candidates expressed strong interest in inspiring more girls to pursue STEM and giving back to the society and scientific community. Though some cited positive personal experiences with a mentor and how that influenced their academic and professional lives, several expressed the gender-specific prejudices faced when growing up.

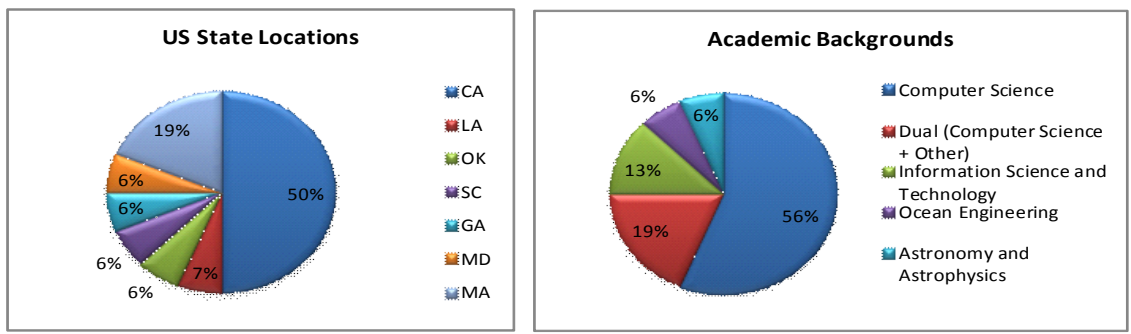

Figure 2. Information on 16 mentors (a) US State Locations (b) Academic Backgrounds

As of March 2013, eight schools (five public, two private, one charter), located in California and Massachusetts, are associated with MAGIC. In addition, a few mentees belong to four other schools that are not officially associated with MAGIC but allow student-level participation. To date, MAGIC has remotely mentored 23 girls from grade 6 through pre-college level, as shown in Figure 3a. All the mentees had goals and expectations before starting the program; their goals could be categorized into three classes:
- (i) General learning: Some girls started the program with open-ended learning goals, e.g., to deepen their understanding of biomedical engineering.
- (ii) Specific-skills: Some mentees were very clear about the sort of skills they wanted to learn, e.g., to learn C++ for a robotics project to be undertaken the following year.
- (iii) College and Career Exploration: Some girls required mentoring to prepare college applications or explore their career interests, e.g., a mentee wanted to learn about environmental engineering major and the possible college choices; another mentee was interested in learning about careers in software development.

The distribution of goals across mentees is shown in Figure 3(b).

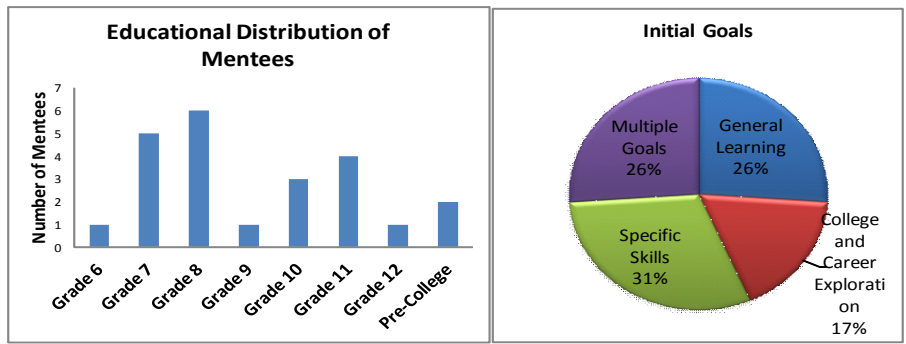

Figure 3. Information on the 23 Mentees (a) School Grade (b) Motivation to join MAGIC

## 5.2 Remote Mentoring Activities

During the first few meetings, the MAGIC pairs experimented with several communication methodologies before deciding to use a combination of tools that work well on the computers at both ends. Figure 4a shows the top tool choices during the past few years. In addition, all the pairs used emails for ad hoc communication; a few had the opportunity to meet with each other in-person once or twice during the course of the mentoring relationship. Figure 4b shows the skills and topics that the mentees learnt through remote mentoring sessions. The majority (42%) of the skills acquired by mentees were related to programming languages such as C and JAVA and programming tools such as Scratch and Alice. Other skills included Web development such as CSS, HTML, JavaScript, creative skills such as creating music through programs, animation in Flash, architecture and modeling, and cryptography codes, and topics such as STEM careers, biomedical engineering, electronic voting, Google maps, Excel, astronomy and space, computer networks, and science experiments.

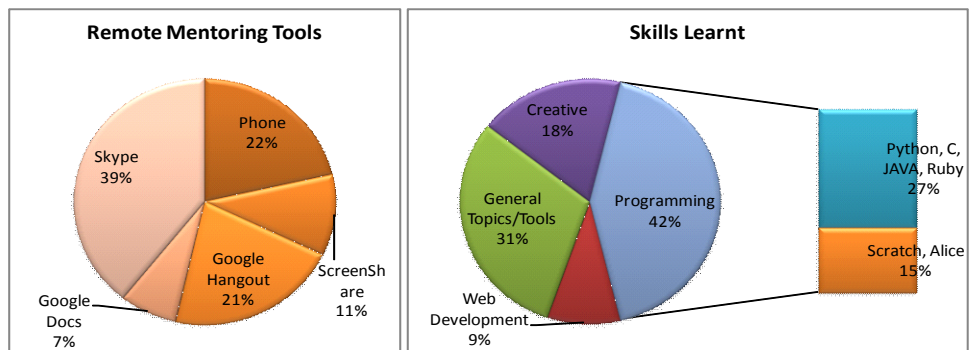

Figure 4. (a) Communication tools used for remote mentoring, (b) Skills learnt through remote mentoring

In addition to several reading assignments, the mentees completed a variety of tangible STEM projects. These projects could be classified into the following categories:

(i) Computation Projects: While learning programming languages, mentees worked on several beginner level programs requiring computations such as calculating factorials, identifying prime numbers, printing Fibonacci series, etc.

(ii) Game/Animation/Web Projects: Several mentees developed projects such as a website with Flash animations, the Tic-tac-toe game, an interactive tool to teach English alphabet to children, etc.

(iii) Non-programming Projects: Several mentees worked on projects such as interviewing and shadowing a STEM professional, preparing college applications, exploring Google SketchUp, and designing science fair experiments, and engaged in outdoor activities such as learning to use a telescope.

Each pair worked on one or multiple projects. The distribution of projects in the above-mentioned categories was 44%, 36%, and 20%, respectively.

## 5.3 Remote Mentoring Challenges

MAGIC mentors and mentees reported certain challenges throughout the remote interactions. Scheduling and time management were challenging for several participants. Many mentees experienced difficulty in assigning time to MAGIC assignments and projects, and reported that they remained busy with several after school activities, full time jobs, and travelling. A few reported time zone difference as a challenge. Some mentors reported that the mentee was involved in too many activities to be able to manage her time efficiently.

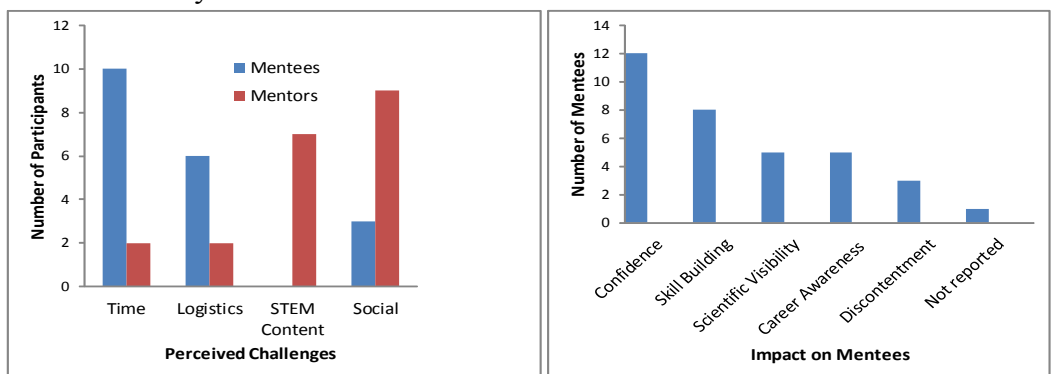

Figure 5 (a) Challenges faced during remote mentoring (b) Impact vs. Number of Mentees

Remote setting logistics were another challenge reported by many, e.g., software version and operating system (Mac vs. Windows) compatibility and lack of infrastructure (webcam, headphones, etc.). Several mentors reported teaching and remote delivery of STEM content as a huge challenge. In a few relationships the mentor did not have the exact skills that the mentee wanted to learn. Moreover, there is a lack of material on teaching advanced programming language to children. Some mentors found it difficult to decide on a particular topic, e.g., Ruby vs. JAVA. Few mentees chose over-ambitious projects for their age, and thus the mentors found it challenging to explain the intricate STEM concepts remotely. Many participants faced social and interpersonal challenges, e.g., lack of family support, miscommunication, presence of international accents, lack of mentee' continued interest, lack of response from the mentee, etc. In addition, some mentors reported difficulty in drawing a line between instructing and mentoring, and in connecting with an introverted mentee. Figure 5a illustrates the type of challenges faced against the number of participants. At least seven mentees and seven mentors reported no challenges throughout the interactions. All the pairs successfully dealt with the challenges on their own or with support from the core team, e.g., the mentee started using a planner in one case and the mentor started creating power point slides to improve communication of complicated topics.

**5.4 Impact of MAGIC Mentoring on Mentors and Mentees**
*Impact on Mentees*: Overall, 12 mentees reported feeling more confident about their skills and personal attributes as a result of the MAGIC interactions. They enjoyed the experience and valued the invested time; eight mentees felt a significant improvement in their STEM skills, resulting from projects such as writing computer programs, developing websites, and learning tools. Some mentees had the opportunity to gain public visibility through the mentoring experience, e.g., six got the opportunity to attend the Grace Hopper Conference for Women in Computing: four as panel members and two as poster presenters; two mentees were winners of the 2013 NCWIT Aspirations Bay Award. A mentee reported that she showcased her website to her family living abroad. At least five mentees reported that they gained significant improvement in career awareness and advancement; one received a reference letter from her mentor for high school admission; two became aware of career choices: space and environmental engineering; one mentee got an internship opportunity as a result of a MAGIC site visit. However, three mentees reported discontent due to the remote aspect. They wished they had more time to devote to MAGIC and complained of their mentors' frequent travels. Also, one mentee did not report any impact due to premature termination. A detailed distribution of the impact on mentees is shown in Figure 5b.

*Impact on Mentors*: The majority of the mentors reported a positive and educational experience and enjoyed their relationship with the young mentees. At least three mentors expressed close interest in mentoring the following year, six mentors offered assistance in MAGIC core and administrative activities such as recruitment, publicizing, and sundry items and expressed interest in becoming board members. However, three mentors did not find the relationship rewarding due to the lack of mentees' commitment and enthusiasm.

## 6. Discussion
### 6.1 Organizational Outcomes
*Accessibility:* Most of our mentors are located in California and Massachusetts because a majority of the MAGIC core and board members reside in these two states. Nevertheless, mentors from five other states have also been recruited. While the mentees are also located in the two headquarter states, MAGIC has created associations with three different types of schools in order to reach out to different societal sections. Even in its formative years, MAGIC has matched mentees and mentors across seven different US states. Interestingly, several pairs are east-coast/west-coast pairs.

*Personalization:* Our mentees not only come from different familial backgrounds, but also have a diverse set of expectations from MAGIC, equally split among skill building, topic-based learning, and career exploration. In the past five years, MAGIC ensured that each mentee is associated with the most appropriate mentor, and offered tailored services to the mentees, wherein they learnt a variety of STEM skills (programming, creative, Web, conceptual) and worked on a variety of projects (technical and interpersonal).

*Plan for Growth:* MAGIC plans to expand its program to tens of schools and hundreds of mentees in the next five years. The biggest challenge would be replicating the energy and dedication of the small board of directors and this is precisely the reason that we have intentionally decided to stay small for the first few years. We plan to grow by partnering with one or two schools every year, or by adding more mentees in a school, or by carefully expanding geographically. Presently, MAGIC has a bank of about 40 mentors, not all of who actively mentor during each session. The MAGIC board has been brainstorming different ways to increase the mentor bank while retaining the organization's extremely high bar for mentors. Some ideas include partnerships with universities, national labs, and companies, possibility of university students getting credit for mentoring hours, shorter mentoring sessions in an effort to reduce the number of hours a mentor needs to commit to, etc.

**6.2 Lessons learnt about Remote Mentoring**

Based on our retrospective analysis, we can draw several conclusions on remote mentoring young females in STEM. The following key findings will help any organization when establishing and implementing the program, recruiting mentors, and above all understanding the general needs of girls aspiring to be STEM graduates and professionals.

- Skype, Google Hangout, and phone are the popular and feasible choices for remote communications endorsed by a majority of our mentor-mentee pairs.
- Programming, topic-based learning, and creative skills (e.g., music, animation software) are the top learning choices of our mentees. These represent the type of skills that could be learnt successfully via remote mentoring. While the big picture goal of MAGIC is to target a variety of STEM fields, the current skill palette is largely dominated by programming languages (Figure 4b) because a majority of our current mentors have strong backgrounds in computer science (Figure 2b). Additionally, the founding members of MAGIC possess similar backgrounds.
- Computational and game projects, and shadowing activities, were the popular project choices among our mentees. These represent the types of finished deliverables that could be accomplished through remote mentoring.
- We find that the mentors and mentees perceived different set of challenges. While the girls were concerned about time management and logistics, the mentors' main concerns were about the delivery of content and social challenges. It is thus important to educate both sides to collectively address the challenges. However, these challenges do not necessarily indicate failures, and were effectively and collaboratively addressed by the MAGIC team, leading to largely positive outcomes.
- Above all, remote mentoring made a positive impact on the mentees' lives, building skills and self-confidence. In addition, several mentees gained scientific visibility and significant career awareness.

**7. Conclusions and Future Work**

A recent Harvard Business Review study[19] suggests that more women in the workforce could raise GDP by 5%. The statistics call for nationwide efforts to fix the problem of the dearth of women in STEM education and workforce. In this paper, we introduce our unique organization, MAGIC, which remotely yet closely engages girls to pursue interests in the STEM fields. MAGIC aims to facilitate highly personalized services to young females, and hence the most pressing concern is recruiting mentors and mentees who have the passion, time, and energy to help realize the vision of MAGIC. Through this study, we have provided a data-driven perception on remote mentoring young females in STEM. One limitation of this study was that a majority of the mentoring projects were focused on computer programming due to the backgrounds of our current mentors. Hence, this study might not have provided a balanced analysis of remote mentoring in the STEM fields in general. To address this, we plan to continue our efforts in recruiting mentors with expertise in other STEM fields. We also plan to conduct the study with increased numbers of pairs, diversify the mentees' locations and interests, and study correlation between the challenges faced and the mentoring outcomes.

## 7. Acknowledgements

We sincerely thank our sponsors (Google, Teradata, and personal donors), participating schools, MAGIC mentors, MAGIC mentees, and mentees' parents. Additionally, we thank the NIH Fellows Editorial Board for their assistance in editing this manuscript.